# Bayesian Conditional GAN for MRI Brain Image Synthesis

Gengyan Zhao[a,*], Mary E. Meyerand[a,b,c], and Rasmus M. Birn[a,d]

[a] Department of Medical Physics, University of Wisconsin – Madison, USA

[b] Department of Radiology, University of Wisconsin – Madison, USA

[c] Department of Biomedical Engineering, University of Wisconsin – Madison, USA

[d] Department of Psychiatry, University of Wisconsin – Madison, USA

[*] Corresponding Author: Gengyan Zhao, Ph.D.,

E-Mail: gzhao23@wisc.edu

# Abstract

As a powerful technique in medical imaging, image synthesis is widely used in applications such as denoising, super resolution and modality transformation etc. Recently, the revival of deep neural networks made immense progress in the field of medical imaging. Although many deep leaning based models have been proposed to improve the image synthesis accuracy, the evaluation of the model uncertainty, which is highly important for medical applications, has been a missing part. In this work, we propose to use Bayesian conditional generative adversarial network (GAN) with concrete dropout to improve image synthesis accuracy. Meanwhile, an uncertainty calibration approach is involved in the whole pipeline to make the uncertainty generated by Bayesian network interpretable. The method is validated with the T1w to T2w MR image translation with a brain tumor dataset of 102 subjects. Compared with the conventional Bayesian neural network with Monte Carlo dropout, results of the proposed method reach a significant lower RMSE with a p-value of 0.0186. Improvement of the calibration of the generated uncertainty by the uncertainty recalibration method is also illustrated.

# Introduction

Image synthesis is an important topic in the field of medical imaging, and many techniques in all levels of medical image processing can be categorized into the field of image synthesis. With image synthesis techniques, MR images can be reconstructed from the data collected in the k-space (Zhu et al., 2018), image denoising can be achieved by generating images with low noise from the images with high noise (Jiang et al., 2018), and the resolution of an image can be improved from a low resolution image, which is also called super-resolution (Sanchez and Vilaplana, 2018). Sparse reconstruction can significantly shorten the scanning time, and freeze the motion for dynamic imaging, while denoising and super-resolution can immensely improve the image quality for diagnosis. With image synthesis techniques, we can even further generate the image of one modality from the image in another modality, or synthesis images across different contrast mechanisms, for example: generating CT images from MR images (Guerreiro et al., 2017; Roy et al., 2014), or synthesizing T2w MR images from T1w MR images (Jog et al., 2017). The former example can reduce the patient's radiation dose and generating CT images when CT scanner is not available, for example in a PET/MR scanner (Liu et al., 2018). The later example can reduce the total scan time of the clinical protocol.

In medical neuroimaging, compared with other modalities, MRI can deliver high resolution, 3D images with a variety of contrast mechanisms in a radiation free manner. Many pulse sequences have been designed to capture different tissue contrast mechanisms for different diagnosis purposes. For example, the pulse sequence of Magnetization Prepared Gradient Echo (MPRAGE) (Deichmann et al., 2000) provides a heavily T1-weighted contrast, which is useful to visualize the cortex and subcortical structures, while the Fluid Attenuated Inversion Recovery (FLAIR) (Hajnal et al., 1992) is a kind of T2w sequence, which is good at catching the white matter lesions in normal white matter and is widely used for imaging multi-sclerosis patients (Simon et al., 2006). Depending on certain diseases and the diagnosis requirements, multiple MR pulse sequences can be acquired during a single MRI session to get a comprehensive information about the brain anatomy and function. In this situation, MR inter-contrast image synthesis has the potential to reduce the total scan time and cost. If an unacquired MR contrast is found to be useful, inter-contrast image synthesis is also able to generate it retrospectively.

Conventional image synthesis approaches can be classified into two categories: registration based image synthesis and intensity transformation based image synthesis. Usually, the problem of image synthesis can be formulated as: given an image $a_1$ in contrast A, synthesizing its corresponding image $b_1$ in contrast B, with an image set, *a*, in contrast A and its corresponding image set, *b*, in contrast B. Usually, the image set *a* and its paired image set *b* are co-registered.

For the registration based image synthesis, the image *a* is registered to $a_1$ with deformable registration algorithms, and then the obtained deformable transformation is applied to *b* to generate $a_1$'s paired image $b_1$ in contrast B. This method was first used to synthesize positron emission tomography (PET) images from MR images as a single atlas registration and transformation approach with a single aligned image pair in set *a* and set *b* (Miller et al., 1993). Then, this method was extended to use multiple aligned image pairs in set *a* and set *b*. All the images in a are registered to $a_1$ with deformable registration. Then the same deformable transformations are applied to the corresponding paired images in *b*, and an intensity fusion is performed to synthesize the intensity of each voxel of $b_1$. This extended method was first used to synthesize the computed tomography (CT) images from MR images for the attenuation correction in PET reconstruction (Burgos et al., 2014, 2013). Obviously, the performance of registration based image synthesis highly depends on the accuracy of the deformable registration, which is challenging for brain images with many fine structures. Moreover, this method cannot be applied to the situations in which the $a_1$ image has abnormal anatomy, for example, brain images with stroke, brain tumors or multiple sclerosis. Since the anatomic structures in *a* and *b* do match with those in $a_1$, the method is not able to generate $a_1$'s corresponding structures in $b_1$ with contrast B (Burgos et al., 2014; Cardoso et al., 2015; Miller et al., 1993).

Intensity transformation based image synthesis can be viewed as a supervised prediction approach. First, for every voxel location in an image in set *a*, a feature vector is extracted. The feature vector, for example, could be a 3D patch centered at that voxel. Each feature vector has a target intensity value, which is the voxel intensity at the same location in the corresponding image in set *b*. The training dataset is created by extracting the feature vectors and generating the feature-vector-target-intensity pairs across all in the images in the image sets *a* and *b*. Then, a regression algorithm can be learned to map the feature vectors to their target intensity values. During the prediction stage, feature vectors can be extracted from the image $a_1$ to generate the corresponding voxel intensities for the image $b_1$ with the learned regression algorithm. Image analogy was one of the earliest intensity transformation models proposed for image synthesis (Hertzmann et al., 2001). The patch extracted from $a_1$ is matched to the its *k* nearest-neighbor patches in the patches extracted from *a*.

Then the *k* corresponding patches extracted from *b* were combined to generate the patch for $b_1$. In MR image synthesis, this method was used for the purpose of image registration (Iglesias et al., 2013). MIMECS is another image synthesis method proposed for intensity transformation, which solves the problem with dictionary learning (Roy et al., 2013, 2011). Patches extracted from *a* are treated as a set of bases of a dictionary, and are used to sparsely represent the patches extracted from $a_1$ with the linear combinations of these bases. Each basis patch extracted from *a* has its corresponding patch extracted from *b*, and the linear combinations of these corresponding patches from *b* with the same weights are used to generate the patches for $b_1$. Intensity transformation based methods are usually computationally intensive, and cannot synthesize all the contrasts flexibly (Roy et al., 2013). For a more comprehensive review, please refer to (Jog, 2016).

As learning based models develop rapidly in the recent past, machine learning and deep learning based techniques were also applied on the task of image synthesis. REPLICA encodes both global and local information in the feature vectors and uses random forests to learn the nonlinear regression for image synthesis (Jog et al., 2017). CNN was used to synthesize the CT images from the corresponding PET images for the attenuation correction in PET reconstruction (Liu et al., 2018). CT images were also synthesized from MR images by CNNs (Xiang et al., 2018) and GANs (Nie et al., 2018) for patient dose reduction and MR-only treatment planning in radiation therapy.

The purpose of this work is to build a deep learning based model to increase the accuracy in image synthesis as well as generating model uncertainty for each synthesized image in MR contrast transformation. In this study, we propose Bayesian conditional GAN with concrete dropout and a model recalibration method to increase the accuracy in image synthesis and improve the calibration of the uncertainty generated.

In comparison to previous image synthesis studies, our study has several novel aspects. First, we propose Bayesian conditional GAN as the main image synthesis engine for this task. As a Bayesian neural network it can not only synthesize the image in the target contrast but also generate the uncertainty map for the synthesized image as well. The uncertainty map can be the source that our judgement on whether the synthesized image can be trusted or not is based on. Second, we use concrete dropout in the Bayesian neural network instead of the conventional Monte Carlo dropout (Gal and Ghahramani, 2016). As a gradient-tuned dropout, the dropout rate of concrete dropout can converge to its optimal value during the training stage. This eliminates the complex grid search procedure to find the best dropout rate for each Monte Carlo dropout layer, and also ends up with a better calibrated uncertainty. Finally, we incorporate a model recalibration method as a post-processing approach in the model to further improve the calibration of the posterior distribution of the predicted voxel values and the corresponding model uncertainties.

The accuracy and robustness of the model were evaluated on the challenging application of T1w to T2w brain tumor image synthesis. We hypothesize that the Bayesian conditional GAN with concrete dropout and model recalibration is suitable for the inter-contrast MR brain image synthesis with accurate predictions and well-calibrated uncertainties, which can reflect the confidence levels of the model on the predictions.

## Material and Methods

### Dataset and Preprocessing

This work used the T1w and T2w MR brain volumes of 102 pre-operative subjects of The Cancer Genome Atlas (TCGA, cancergenome.nih.gov) Glioblastoma Multiforme (GBM) collection. The data were released by the international multimodal BRAin Tumor Segmentation challenge (BRATS 2018, https://www.med.upenn.edu/sbia/brats2018/) (Menze et al., 2015) through The Cancer Imaging Archive (TCIA, www.cancerimagingarchive.net). The brain images were collected from 8 institutions with 3T scanners of different vendors and having different MR imaging sequence implementations. The data were distributed after preprocessing. All the brain volumes were co-registered to the same anatomical template with affine registration, and then were resampled into the same 1 mm$^3$ resolution. Finally, all the brain volumes were skull-stripped. Detailed patient information, scanner information and imaging information for each image can be found in (Bakas et al., 2017).

**Bayesian Conditional GAN**

Conditional GAN (Isola et al., 2016) is an accurate and consistent approach to synthesize images. To make it also have the ability of generating model uncertainty for each prediction, we propose to convert it into a Bayesian neural network. In the framework of Bayesian deep learning, all the variables in the neural network and the predicted results are treated as random variables following certain distributions. The training purpose of Bayesian deep learning is to estimate the posterior distribution $p(\omega|X,Y)$ of the weights in the neural network, which is usually intractable. Thus, a weight distribution $q_\theta(\omega)$ with the a parameter set $\theta$ is used to approximate the intractable posterior, and the training procedure is equivalent to minimizing the KL divergence between them (Gal and Ghahramani, 2015).

$$\mathcal{L}(\theta) = KL(q_\theta(\omega) \| p(\omega|X,Y)) \quad (0.1)$$

With the techniques in variational inference and Monte Carlo integration, the KL divergence in Eq. 4.1 can be simplified as the following loss function (Gal and Ghahramani, 2015):

$$\mathcal{L}(\theta) = -\frac{N}{M}\sum_{i \in S} \log p(y_i | f^{g(\theta,\epsilon)}(x_i)) + KL(q_\theta(\omega) \| p(\omega)) \quad (0.2)$$

where $N$ is the total number of observations and $M$ is the number of observations used in the current training step. $g(\theta,\epsilon)$ is the collection of the weight random variables in the Bayesian neural network with the collection of the weight matrices, $\theta$, and the collection of the random variables, $\epsilon$, which follow Bernoulli distributions. $x_i$ is the $i$th input data and $y_i$ is the ground truth of the $i$th input data. $f^{g(\theta,\epsilon)}(x_i)$ is the predicted result from the forward pass of the Bayesian neural network. It can be proved that the training of a Bayesian neural network is equivalent to the training of a conventional neural network with the dropout regularization and the square of the $l^2$ norm regularization of the weight matrices in the conventional neural network (Gal and Ghahramani, 2015). Therefore, by plugging in dropout layers, the conventional conditional GAN can be changed into a Bayesian conditional GAN.

During the testing stage, unlike the dropout layers in a conventional neural network, the dropout layers in a Bayesian neural network will still function in the forward passes. This procedure is called dropout testing, and it is equivalent to sampling the posterior distribution of the predicted random variables (Gal and Ghahramani, 2016).

$$p(y^* | x^*, X, Y) \approx \int p(y^* | x^*, \omega) q_\theta^*(\omega) d\omega =: q_\theta^*(y^* | x^*) \tag{0.3}$$

Given the assumption that these predicted random variables follow normal distributions, we can use the mean and the variance of the predicted values from multiple forward passes as the unbiased estimators of the mean and variance of the distributions of the predicted random variables. The mean can be used as the final prediction result of the input data, and variance can be used as the model uncertainty for the prediction (Gal and Ghahramani, 2016).

**Concrete dropout**

One drawback of the conventional Monte Carlo dropout is that the dropout probability is a hyperparameter, and need to be tuned manually. Grid-searching over the entire space of dropout probabilities for all the dropout layer is exhausting work and will cost an immense amount of computation power. Moreover, the predicted posterior distribution $p(y^* | x^*, X, Y)$ in dropout testing can be affected by the dropout probability. Thus, the accuracy of the final predicted result and the calibration of the model uncertainty can also be greatly influenced by the hand-tuned dropout probability.

Given the assumption that the prior of the weights, $p(\omega)$, also follows normal distribution, the KL divergence in the simplified loss function, Eq. 4.2, can be proved to be proportional to the following equation (Gal et al., 2017):

$$KL(q_\theta(\omega) \| p(\omega)) \propto \sum_{DropoutLayers} \frac{1}{2} l^2 (1-p) \|M\|_2^2 - K\mathcal{H}(p) \tag{0.4}$$

where the sum is calculated over all the dropout layers; $l$ is a hyperparameter; $p$ is the dropout probability; $M$ is the weight matrix in the layer before the dropout layer; $K$ is the number of input channels for the dropout layer; $\mathcal{H}(p)$ is the entropy of a Bernoulli random variable with the probability $p$:

$$\mathcal{H}(p) := -p \log p - (1-p) \log(1-p) \tag{0.5}$$

The entropy term in the regularization only depends on the dropout probability, which means it will be completely ignored when the weights of the network are the only variables to be optimized. However, it enables a gradient-tuned dropout probability when the dropout probability is part of the optimization variables. This means the dropout probability will converge to its optimal value as the training proceeds. This spares the effort of a grid-search for the optimal dropout probability for each dropout layer, and will also result in a better calibration of the predicted posterior and the model uncertainty. The whole structure of a Bayesian conditional GAN with concrete dropout is illustrated in Fig. 1.

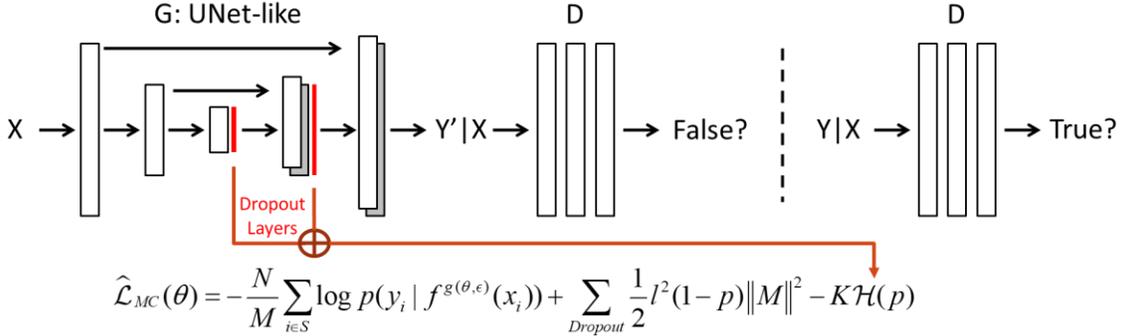

Fig. 1. Illustration of the structure of a Bayesian conditional GAN.

Since the derivative of the KL divergence term, $KL(q_\theta(\omega) \| p(\omega))$, need to be calculated with respect to $p$ during the back propagation, we use the concrete distribution relaxation of dropout's discrete Bernoulli distribution, which reparametrizes the distribution as the following (Gal et al., 2017):

$$\tilde{z} = sigmoid(\frac{1}{t}(\log p - \log(1-p) + \log(u) - \log(1-u))) \tag{0.6}$$

where $t$ is the hyperparameter, temperature, and $u \sim \mathcal{U}(0,1)$. This concrete relaxation of the dropout operation is referred to as concrete dropout (Gal et al., 2017). With it we can optimize the dropout probability in the training stage of a Bayesian neural network.

**Model Recalibration**

The image synthesis in this work is achieved by the proposed Bayesian conditional GAN, and the predicted posterior and the corresponding model uncertainty are generated by the concrete dropout. However, as a variation inference technique, Bayesian deep learning cannot guarantee the absolute accuracy of the predicted posterior and the model uncertainty. Thus, to further improve the accuracy of the predicted posterior and the model uncertainty, we propose to incorporate a model recalibration procedure for Bayesian deep learning.

Based on its probabilistic definition in Bayesian statistics, a 95% credible interval should be able to catch the value of interest (e.g. ground truth) with a 95% probability. With probabilistic calibration and a calibration dataset, the predicted posterior can be mapped to the true distribution, which can accurately reflect the probabilistic definition of credible interval. In the Bayesian deep learning model for image synthesis, for each voxel $t$ the model generates a posterior distribution, a probability density function (PDF), targeting the ground truth value $y_t$ during the testing stage. The PDF can be converted to a cumulative distribution function (CDF), $F_t$, which can be used to perform the model's probabilistic recalibration (Kuleshov et al., 2018):

$$p \to f = \frac{\sum_{t=1}^{T} \mathbb{I}\{y_t \leq F_t^{-1}(p)\}}{T} \text{ for all } p \in [0,1] \quad (0.7)$$

where $F_t^{-1}(p) := \inf\{y : p \leq F_t(y)\}$ is the quantile function. From Eq. 4.7 it can be seen that the model recalibration procedure maps the probability of the CDF of the predicted posterior to the real probability that the ground truth values of the voxels in the calibration dataset fall in the corresponding credible interval. This exactly follows the probabilistic definition of credible interval. The model recalibration procedure should calibrate all the $p$ values for the Bayesian deep learning model according to the calibration dataset after the training of the Bayesian deep learning model. Then, during the testing time the predicted posteriors from the Bayesian deep learning model can be mapped to the calibrated posteriors.

**Experiments**

Before doing image synthesis, each subject's original 3D brain volume was dissembled along the longitudinal (superior-inferior) axis into a stack of 2D images, and then each 2D image was normalized to [0,1]. Before sent to Bayesian conditional GAN, all the T1w images and T2w images were resampled to the size of 286×286 with bilinear interpolation. Since all the subjects have brain tumors at different locations with different sizes, a 256×256 window was randomly shifted within the 286×286 brain image to cut a 256×256 image to send to the Bayesian conditional GAN at each iteration as a data augmentation approach. The Bayesian conditional GAN used a UNet-like (Ronneberger et al., 2015) convolutional encoder-decoder network as the generator and a CNN with 5 convolutional layers as the discriminator. Concrete dropout layers were plugged into the network structure after the $2^{nd}$, $3^{rd}$ and $4^{th}$ transposed convolutional layers. Batch normalization was used in the network, and a batch size of 16 was used during the training stage. A combination of the conditional GAN loss, the $l_1$ norm between the synthesized image and the ground truth, and the regularization of the KL divergence term, $KL(q_\theta(\omega) \| p(\omega))$, Eq. 4.4, was used as the loss function. A weight of 100 was used as the weights for both the $l_1$ norm and the KL divergence term in the loss function. Within the KL divergence term a weight of 1e-6 was used for the network weight regularization term and a weight of 1e-5 was used for the concrete dropout regularization term. The hyperparameter, temperature, $t$, in the concrete distribution relaxation was set as 0.1. The Bayesian conditional GAN was trained by the Adaptive Moment Estimation (ADAM) (Kingma and Ba, 2014) algorithm with a fixed learning rate of 0.0002 and the momentum: $\beta_1 = 0.5$ and $\beta_2 = 0.999$. After 40 epochs of training, there was no obvious improvement of the loss, and the network weights from the $40^{th}$ epoch was used for the following tests. After prediction, all the synthesized images and the ground truth images were normalized to the range of [0,255] for later visualization and result analysis.

The whole processing pipeline was implemented on the platform of PyTorch (Paszke et al., 2017) based on the original work of conditional GAN (Isola et al., 2016). All the training and testing of the proposed method and the evaluation of other compared methods was performed on a workstation hosting 2 Intel Xeon(R) E5-2620 v4 CPUs (8 cores, 16 threads @2.10GHz) with 64 GB DDR4 RAM and an Nvidia TITAN Xp GPU with 12 GB GPU memory. The workstation runs a 64-bit Linux operation system.

In the 102 subjects, 82 subjects were used for training and 20 subjects were used for testing. The image synthesis performances and the generated model uncertainties of the Bayesian conditional GANs with the concrete dropout and the conventional Monte Carlo dropout were compared. For Monte Carlo dropout a dropout rate of 0.5 was used. In the neural network structure, the positions of the Monte Carlo dropout layers were the same as those of the concrete dropout layers, but no KL divergence term was used in the loss function for the Monte Carlo dropout. In the model recalibration procedure, the training dataset was used for the model recalibration. No obvious overfitting was observed.

## Results

**Image Synthesis: Prediction Accuracy and Model Uncertainty**

The image synthesis accuracy of the proposed Bayesian conditional GAN with concrete dropout was evaluated and compared with that of the Bayesian conditional GAN with Monte Carlo dropout. In Fig. 2 the boxplots of the root mean square (RMS) errors of each subject's synthesized brain volumes were shown. Overall, the brain volumes synthesized with concrete dropout is more accurate than those with Monte Carlo dropout. A two-sided paired t-test was performed to compare the performance of the two methods, and the p-value of 0.0186 shows that the synthesized brain images with concrete dropout are significantly more accurate than the ones synthesized with Monte Carlo dropout at the 0.05 significance level.

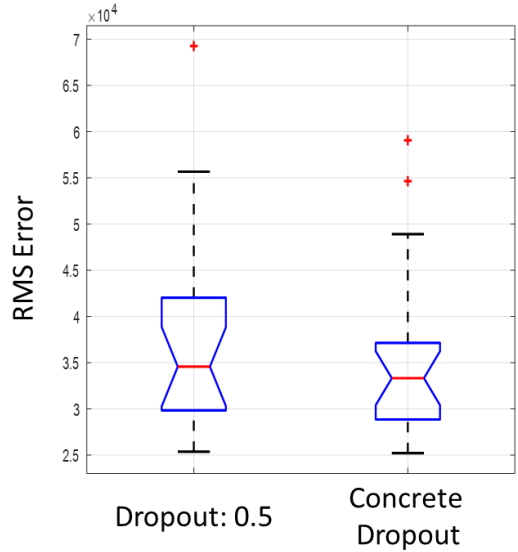

Fig. 2. Accuracy of the synthesized images. The boxplots of the RMS errors of the synthesized brain volumes for the subjects in the testing dataset are shown. The accuracy of the synthesized images by Bayesian conditional GAN with Monte Carlo dropout and concrete dropout were compared. The dropout rate of the Monte Carlo dropout was set as 0.5. In this figure, the points with red '+' symbols are drawn as outliers, if they are greater than q3+1.5(q3-q1) or less than q1-1.5(q3-q1), where q1 and q3 are the first and third quartiles respectively.

Fig. 3 shows the image synthesis results of a representative subject at 3 different slices. The original T1w image, the ground truth T2w image, the synthesized T2w image, the absolute error map between the prediction and the ground truth, and the uncertainty map for both methods are shown. In general, the synthesized T2w images can accurately reflect the brain anatomic structures in the real T2w images. However, there are also regions with relatively large image synthesis errors. For example, in slice 1 the upper left edge of the brain and the lower right spot in the brain marked by the green arrows in the absolute error maps show relatively large synthesis errors, and the uncertainty maps generated by both methods catch these regions with large distribution standard deviations or uncertainties, which are also marked by the green arrows. In slice 2, it can be observed that over all the uncertainty maps generated by both methods can catch the regions with large errors in the absolute error maps. However, the uncertainty map generated by Monte Carlo dropout has a hot spot around the upper left edge of the brain, but the error in corresponding region in the error map is not very large. In addition, if we look into the details of the absolute error maps and the uncertainty maps of both methods, the model uncertainty of each voxel may not always be directly proportional to the corresponding synthesis error in the error map. This is mainly because by definition higher model uncertainty only means that the model is not very confident about the prediction and there is a higher chance for the model to make a prediction with a large error, but this does not mean that the model will definitely make a prediction with a large error at this voxel. In slice 3, we can see that both methods' uncertainty maps catch the high error region in the center of the tumor, but miss the high error region around the edge of the tumor. The possible reason for this is that the brain tumor of every subject in the training dataset has a different shape, location and contrast, and it's challenging for the model to catch the consistency among them.

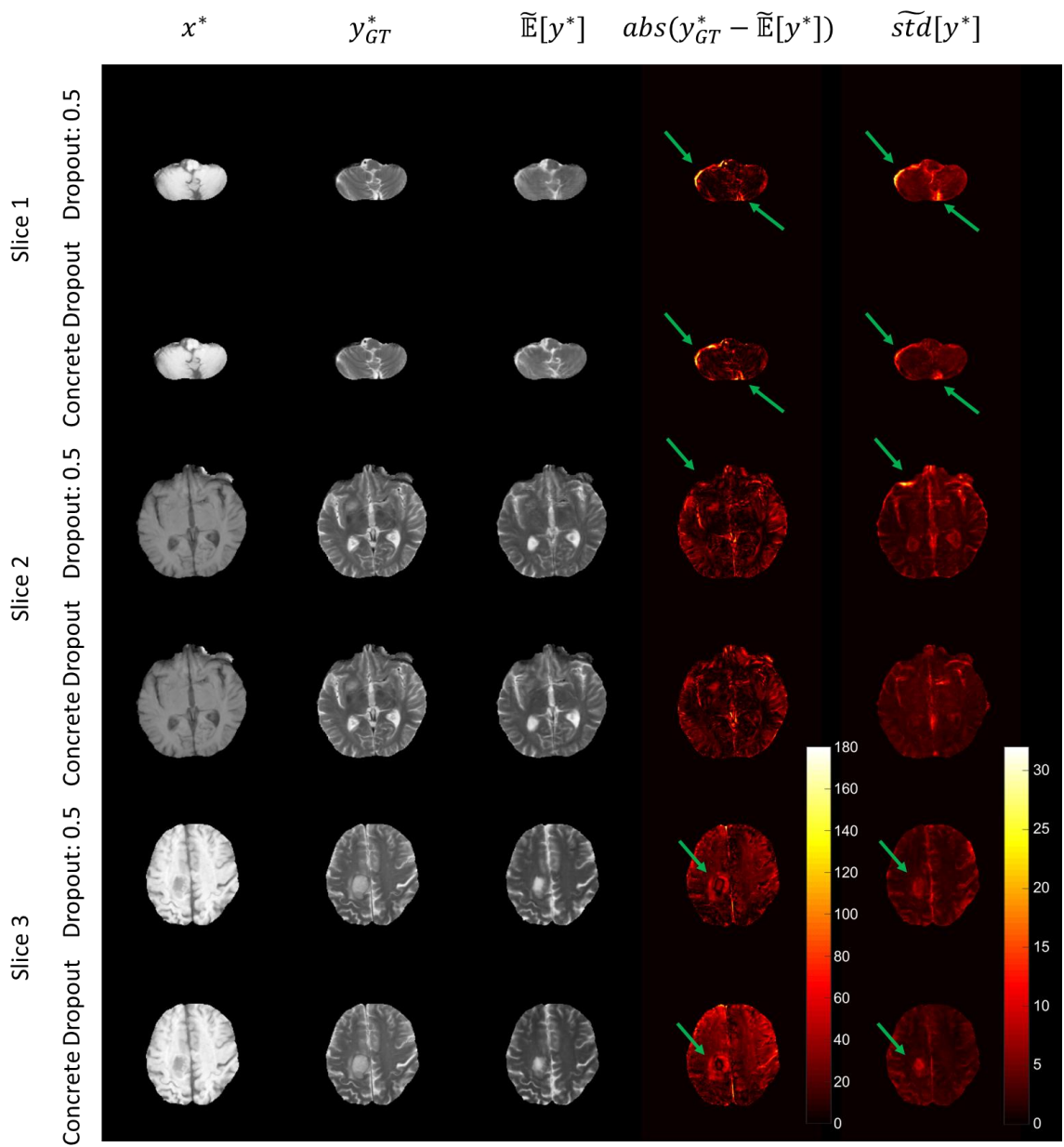

Fig. 3. Image synthesis results of a representative subject at 3 different slices. The original T1w image, the ground truth T2w image, the synthesized T2w image, the absolute error map between the prediction and the ground truth and the uncertainty map for both methods are shown.

**4.3.2 Relationship between Prediction Accuracy and Model Uncertainty**

By definition a voxel with high model uncertainty only means the model is not confident about the prediction on it, and the voxel has a higher chance to get a large prediction error. With this definition, we cannot relate model uncertainty to prediction accuracy directly, but their relation should become stronger with large number of predictions and dimension reduction calculations, e.g. sum or average. Therefore, we plot out the accuracy versus uncertainty relations at 3 different levels in Fig. 4. For the voxel level, we plot the absolute error between the ground truth voxel value and the predicted value against the standard deviation (std) of the predicted posterior distribution. Since for each 3D brain volume the number of voxels in the brain region is huge, we only plot out the data in the brain region for a representative subject. For the slice level and the volume level, we plot the normalized root mean square error (nRMSE) of the predicted slice or 3D volume respectively against the normalized std (nSTD) of that slice or 3D volume. The definition of nRMSE and nSTD are:

$$nRMSE = \frac{\|y - \hat{y}\|_{2,\phi}}{\|y\|_{2,\phi}} \tag{0.8}$$

$$nSTD = \frac{\|std\|_{2,\phi}}{\sqrt{N_\phi}} \tag{0.9}$$

where $y$ and $\hat{y}$ are the ground truth and the predicted value for a single voxel respectively; $\|\cdot\|_{2,\phi}$ denotes the $l^2$ norm over the brain region $\phi$ in that slice or 3D volume; $std$ is the standard deviation of the predicted posterior distribution of a single voxel, and it is used as the model uncertainty for that voxel; $N_\phi$ is the total number of voxels in the brain region $\phi$ in that slice or 3D volume. From Fig. 4 we can find that, as the analysis goes to a more summary level, the relation between the prediction accuracy and the generated uncertainty becomes stronger. This holds true for both methods. This means that as a clue for the prediction

error, the uncertainty of a voxel may not work very well, but the uncertainty of a subject's 3D image volume has much stronger proportional relation with the prediction error.

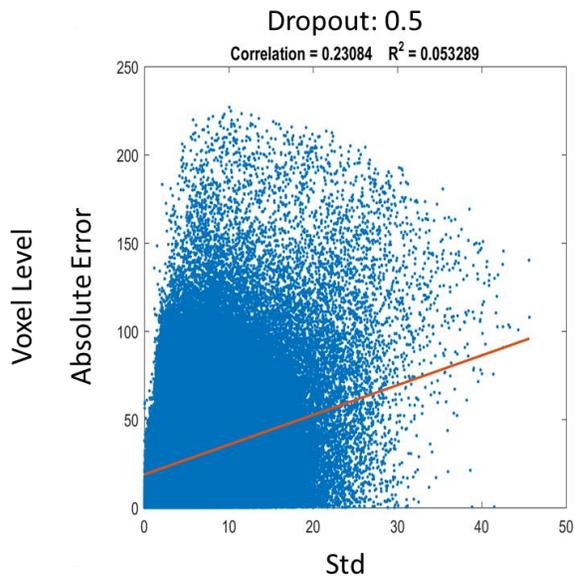 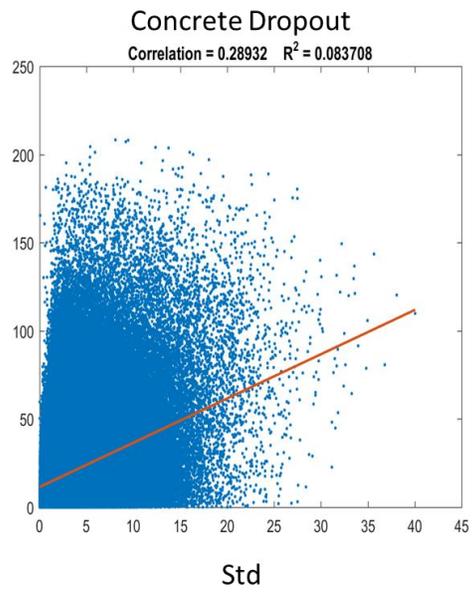

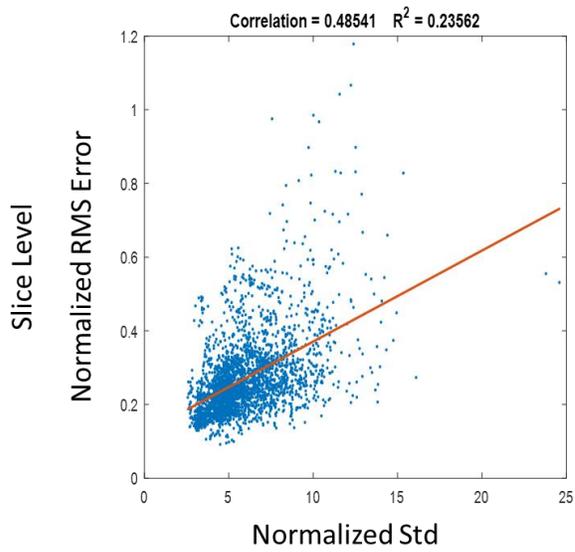 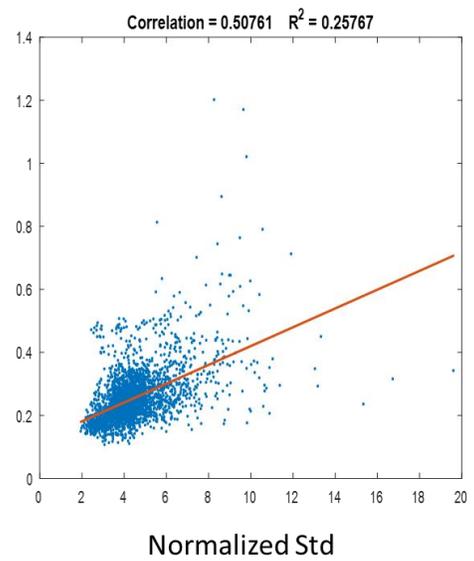

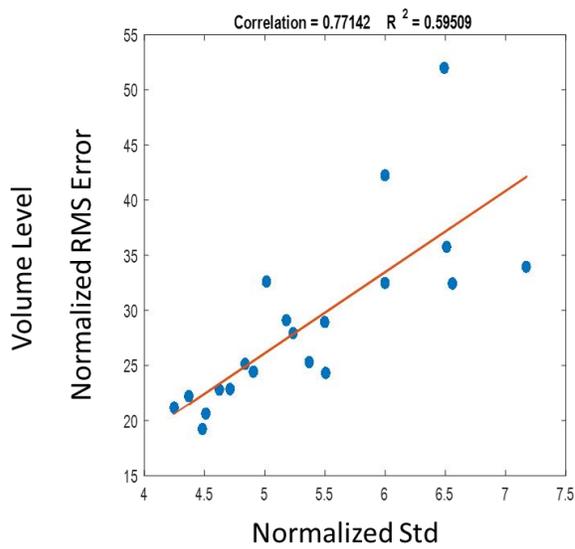 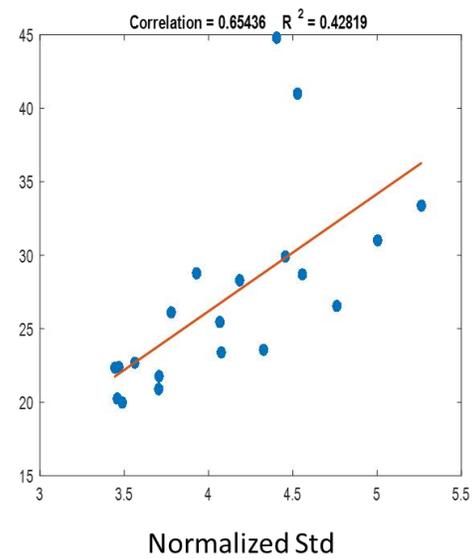

Fig. 4. Relationship between the prediction accuracy and the model uncertainty at different levels. For both methods the error VS uncertainty observations were plotted at the voxel level, the slice level and the volume level. For the voxel level, the absolute error and the sdt of the predicted posterior are used. For the slice level and the volume level, the nRMSE and the normalized std are used. The voxel level plots only include all the voxels from a representative subject. The slice level plots and the volume level plots include all the slices and volumes in the tesing dataset respectively.

**Model Uncertainty Evaluation**

Since the model uncertainty generated is not necessarily directly proportional to the prediction error by its definition. The better metrics to evaluate the uncertainty generated are the precision recall plot and the uncertainty calibration plot. Fig. 5 is the precision recall plot for both methods, which shows how the prediction RMS error of the predictive model changes as the voxels with uncertainty larger than various percentile thresholds are removed. For example, on the recall axis, the point 1 means that all the voxels in the testing dataset are taken into account when the RMS error is calculated, and 0.9 means the RMS error is calculated without the top 10% voxels with the largest model uncertainty values in the testing dataset. First, as we can see, for both curves the RMS error decreases as voxels with uncertainties larger than various percentile thresholds are removed from the RMS error calculation. Since both curves monotonically decreasing as voxels with relatively large uncertainties are removed, this means that the RMS error of voxels with smaller uncertainties are also smaller, and that for both methods the RMS error correlates with the model uncertainty generated with a large number of predictions and the dimension reduction calculation, RMS error. In addition, both curves decrease faster, and the absolute values of the gradients become larger, as the same number of voxels with higher model uncertainties are removed. Second, at each recall value – each model uncertainty percentile threshold – the prediction RMS error achieved by the concrete dropout is smaller than that by the Monte Carlo dropout. This means that the model of Bayesian conditional GAN with concrete dropout is more accurate than the model with Monte Carlo dropout in this application at each model uncertainty percentile threshold.

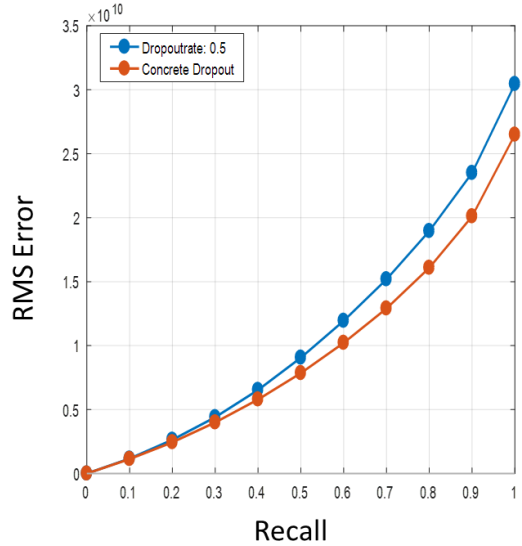

Fig. 5. Precision recall plot for Bayesian conditional GANs with Monte Carlo dropout and concrete dropout.

To analyze the quality of the uncertainty generated by the Bayesian conditional GANs, we studied the uncertainty calibration plot of the models on the testing data. To generate the uncertainty calibration plot, we select a number of probabilities with equal intervals in the CDF of the predicted posterior for each voxel. Then the frequency of the ground truth values of all the voxels falling below the corresponding quantiles of each selected probability is treated as the true probability. The definition of the selected probability and the true probability is the same as the $p$ and $f$ in Eq. 4.7. Then, we can plot out the true probabilities against the selected probabilities as the uncertainty calibration plot. The uncertainty generated with better quality should be closer to the diagonal line, $y = x$. Fig. 6 shows the uncertainty calibration plots for Bayesian conditional GANs with Monte Carlo dropout and concrete dropout. As we can see the uncertainty calibration plots from both methods have error in their scales, and this can be corrected by the model recalibration procedure later. However, the uncertainty calibration plot of the Monte Carlo dropout misses the central point (0.5, 0.5), which is marked by a green point in the figure, and the whole curve is not symmetric to the central point. This means that compared to the true posterior distribution, the posterior predicted by the Monte Carlo dropout is shifted and distorted. First, by distortion, the predicted posterior doesn't align with the normal distribution assumption well. Second, the median of the predicted posterior is not the median of the true distribution. If we use the median, which is the same value as the mean in a normal distribution, as the final prediction result, the prediction will end up with a large prediction error. In contrast, the uncertainty calibration plot of the concrete dropout catches the central point and is symmetric to the central point. Although it also has an error in the scale, this can be corrected with the model recalibration procedure.

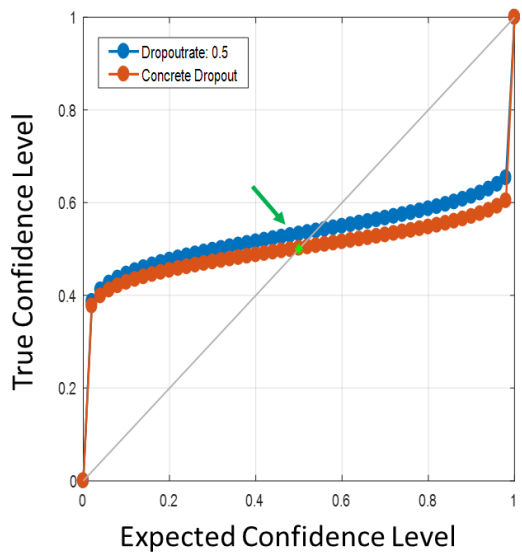

Fig. 6. Uncertainty calibration plot for Bayesian conditional GANs with Monte Carlo dropout and concrete dropout. A perfect calibration of the uncertainty corresponds to the diagonal line, $y = x$, shown in gray in the figure. The central point (0.5, 0.5) is marked as a green asterisk. The expected confindence level is the probability in the predicted posterior distribution, and the true confindence level is the oberserved true frequency in the data.

**<u>Model Recalibration</u>**

To further improve the quality of the uncertainty generated by the models we incorporated the model recalibration procedure after training the Bayesian conditional GANs. Instead of using a separate calibration dataset, we used the training dataset for model recalibration, and no obvious overfitting was observed. After training, predictions were made on the training dataset. With the ground truth of the training dataset the relationship between the selected probability $p$ in the predicted posterior and the true probability, the observed frequency, $f$, can be calibrated according to Eq. 4.7. Then, in the prediction on the testing dataset, the probabilities in the predicted posterior was mapped to the true probabilities according to the calibrated relationship between them. In Fig. 7, the uncertainty calibration plots before and after the model recalibration procedure are illustrated for both methods. As we can see, for both methods the uncertainty calibration plots are closer to the diagonal line after model recalibration. This means that the model recalibration approach can improve the accuracy of the predicted posterior as well as the generated model uncertainty. By calculating the RMS error between the uncertainty calibration plot after model recalibration and the *y* = *x* line, we can see that after model recalibration the calibration error of concrete dropout (RMS error = 0.2868) is still smaller than that of Monte Carlo dropout (RMS error = 0.3783).

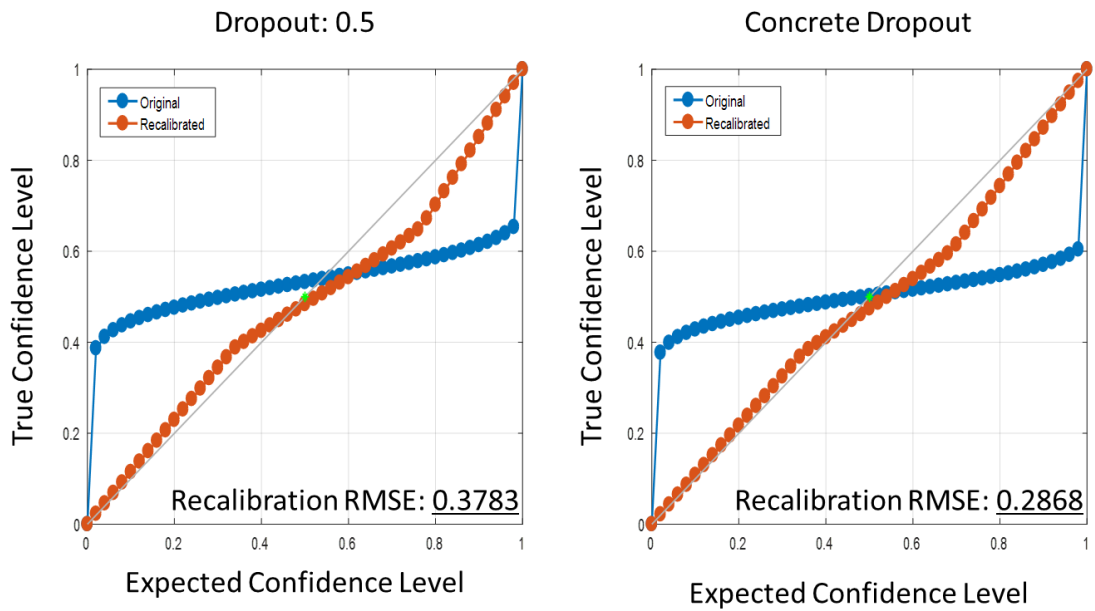

Fig. 7. Uncertainty calibration plot for Bayesian conditional GANs with Monte Carlo dropout and concrete dropout, before and after model recalibration. The point (0.5, 0.5) is marked as a green asterisk. The expected confindence level is the probability in the predicted distribution, and the true confindence level is the oberserved true frequency in the data.

## Discussion

A new image synthesis model is proposed using Bayesian conditional GAN as the main image synthesis engine for the challenging application of MR brain tumor image synthesis. With the framework of Bayesian deep learning the proposed method can generate the posterior distribution of each voxel value for the synthesized image, which gives it the ability to make predictions as well as generate model uncertainties for the predictions. The use of concrete dropout enables the gradient-tuned dropout probability and results in more accurate prediction and uncertainty. By involving the model recalibration approach the calibration quality of the predicted posterior and the generated model uncertainty is further improved. The proposed method was applied to the challenging task of MR brain tumor image synthesis. In comparison with the model with the conventional Monte Carlo dropout, the superior performance of the proposed method was validated.

As we can see in Fig. 2 and 5, the prediction accuracy of the Bayesian conditional GAN with concrete dropout is significantly better than that with Monte Carlo dropout. The possible reason is that the concrete dropout technique offers each dropout layer a gradient-tuned dropout probability which can converge to its optimal value during the training of the network. This optimal dropout probability can help the Bayesian deep learning model reach a higher accuracy than the dropout rate set by experience in the Monte Carlo dropout. At the same time, the concrete dropout changes the hyperparameter, dropout rate, in the forward model to a variable in the loss function for training and saves the time for grid-searching the optimal value of dropout rate. This also makes the model more robust than the model with empirical dropout rate.

In more details, the optimal dropout probability reached by concrete dropout can make the Bayesian deep learning model generate more accurate posterior distributions during the testing time, which is shown in Fig. 6, and that is why it can end up with a higher prediction accuracy. In Fig. 6, the error in the scale of calibration plot is not important, since we can still get a correct prediction when using the mean of the posterior as the final prediction. When the std of the posterior is used as the uncertainty, the uncertainty is only affected by a scale value. However, the shift or distortion of the calibration plot will end up with a wrong final prediction if the mean or median of the posterior is used as the final prediction result, since the final prediction result calculated is no more the median in the true distribution. The distortion of the calibration plot will also make the std uncertainty inaccurate and hard to correct. In Fig. 7, we can see that the error in the scale of the calibration plot can be easily corrected with the model recalibration procedure. The shift and distortion in the calibration plot can also be corrected to some extent, but still will result in a slightly larger RMS error in the calibration plot after model recalibration.

The current workflow of artificial intelligence based research in the field of medical imaging looks like this: a deep learning model is trained on a training dataset, and then it will be applied to a testing dataset to make predictions. The prediction results will be compared with the ground truth given by radiologists. Obviously, various kinds of inconsistency will cause errors in the prediction: e.g. the inconsistency within the training dataset, the inconsistency between the training dataset and the testing dataset, etc. Therefore, if a trained AI is distributed to different hospitals and works by itself in the daily clinical routine, we won't have the ground truth any more, and we won't know whether the predictions made by the AI can be trusted or not. Bayesian deep learning based models solve this problem by generating model uncertainty information for each prediction. Whenever the model uncertainty is above a certain threshold, human intervention can be started to double check the case, or more information of the patient can be required for the AI to make a more confident prediction. By its definition, the model uncertainty may not directly reflect whether a prediction has a small error or a large error, but it suggests the possibility of a small or large error in the prediction. This means by having a large number of predictions and with the dimension reduction calculation we can get an averaged uncertainty having stronger correlation with the prediction accuracy. In Fig. 3 and 4, it is shown that at the voxel level the model uncertainty may not be directly proportional to the prediction absolute error, which will make it harder for us to locate the voxels with possible large prediction errors. However, at the slice level and the volume level the proportional relationship is much stronger, we can at least locate the slice or the brain volume that likely have a large prediction error.

There are also some limitations in this study. To have a more complete understanding of the performance we still need to compare our methods with many conventional image synthesis models. Another limitation is that we only applied our method to the application of MR brain tumor image synthesis from T1w to T2w images. The performance of the method can be different on different applications, so more experiments are still needed to verify the performance and the characteristics of the proposed method. In addition, by including more information in the input data, deep learning based methods have a higher chance to generate more accurate results. For example, instead of using only T1w MR images to synthesize T2w images. T1w and T2-FLAIR images can be used together to synthesize T2w images, or T1w, T2w and PDw images can be used together to synthesize T2-FLAIR images. Since more information is offered, and T1, T2 and proton density are the 3 basic tissue properties for MR tissue contrasts, the image synthesis accuracy achieved by deep neural network can be higher. These studies will be included in our future work.

## Conclusion

This study presents a new Bayesian deep learning based image synthesis model, Bayesian conditional GAN, which can not only accurately synthesize MR neuroimages but also generate uncertainty maps for the synthesized images. The model takes advantage of the concrete relaxation of the Bernoulli distribution and the KL divergence regularization term in the loss function of Bayesian deep learning for gradient-tuned dropout probabilities, and ends up with higher image synthesis accuracy and more accurate model uncertainty. Moreover, the incorporation of the model recalibration method further improves the model uncertainty calibration. The successful application of the proposed method to the MR brain tumor image synthesis suggests that the method can be further applied to other fields of medical image synthesis.

## Acknowledgement

We gratefully acknowledge the international multimodal BRAin Tumor Segmentation challenge (BRATS) for the data collection, data processing and data sharing. We also acknowledge the support of NVIDIA Corporation with the donation of the Titan Xp GPU used for this research. This work was supported by grants from the National Institutes of Health: U01-NS093650. The content is solely the responsibility of the authors and does not necessarily represent the official views of the National Institutes of Health. This work was done when the author Gengyan Zhao worked as a PhD student and research assistant at University of Wisconsin – Madison.

Ç., Durst, C.R., Dojat, M., Doyle, S., Festa, J., Forbes, F., Geremia, E., Glocker, B., Golland, P., Guo, X., Hamamci, A., Iftekharuddin, K.M., Jena, R., John, N.M., Konukoglu, E., Lashkari, D., Mariz, J.A., Meier, R., Pereira, S., Precup, D., Price, S.J., Raviv, T.R., Reza, S.M.S., Ryan, M., Sarikaya, D., Schwartz, L., Shin, H.-C., Shotton, J., Silva, C.A., Sousa, N., Subbanna, N.K., Szekely, G., Taylor, T.J., Thomas, O.M., Tustison, N.J., Unal, G., Vasseur, F., Wintermark, M., Ye, D.H., Zhao, L., Zhao, B., Zikic, D., Prastawa, M., Reyes, M., Van Leemput, K., 2015. The Multimodal Brain Tumor Image Segmentation Benchmark (BRATS). IEEE Trans. Med. Imaging 34, 1993–2024. https://doi.org/10.1109/TMI.2014.2377694

Miller, M.I., Christensen, G.E., Amit, Y., Grenander, U., 1993. Mathematical textbook of deformable neuroanatomies. Proc. Natl. Acad. Sci. U. S. A. 90, 11944–11948.

Nie, D., Trullo, R., Lian, J., Wang, L., Petitjean, C., Ruan, S., Wang, Q., Shen, D., 2018. Medical Image Synthesis with Deep Convolutional Adversarial Networks. IEEE Trans. Biomed. Eng. 65, 2720–2730. https://doi.org/10.1109/TBME.2018.2814538

Paszke, A., Gross, S., Chintala, S., Chanan, G., Yang, E., DeVito, Z., Lin, Z., Desmaison, A., Antiga, L., Lerer, A., 2017. Automatic differentiation in PyTorch.

Ronneberger, O., Fischer, P., Brox, T., 2015. U-Net: Convolutional Networks for Biomedical Image Segmentation. ArXiv150504597 Cs.

Roy, S., Carass, A., Jog, A., Prince, J.L., Lee, J., 2014. MR to CT Registration of Brains using Image Synthesis. Proc. SPIE 9034. https://doi.org/10.1117/12.2043954

Roy, S., Carass, A., Prince, J., 2011. A compressed sensing approach for MR tissue contrast synthesis. Inf. Process. Med. Imaging Proc. Conf. 22, 371–383.

Roy, S., Carass, A., Prince, J.L., 2013. Magnetic Resonance Image Example-Based Contrast Synthesis. IEEE Trans. Med. Imaging 32, 2348–2363. https://doi.org/10.1109/TMI.2013.2282126

Sanchez, I., Vilaplana, V., 2018. Brain MRI super-resolution using 3D generative adversarial networks.

Simon, J.H., Li, D., Traboulsee, A., Coyle, P.K., Arnold, D.L., Barkhof, F., Frank, J.A., Grossman, R., Paty, D.W., Radue, E.W., Wolinsky, J.S., 2006. Standardized MR imaging protocol for multiple sclerosis: Consortium of MS Centers consensus guidelines. AJNR Am. J. Neuroradiol. 27, 455–461.

Xiang, L., Wang, Q., Nie, D., Zhang, L., Jin, X., Qiao, Y., Shen, D., 2018. Deep embedding convolutional neural network for synthesizing CT image from T1-Weighted MR image. Med. Image Anal. 47, 31–44. https://doi.org/10.1016/j.media.2018.03.011

Zhu, B., Liu, J.Z., Cauley, S.F., Rosen, B.R., Rosen, M.S., 2018. Image reconstruction by domain-transform manifold learning. Nature 555, 487–492. https://doi.org/10.1038/nature25988